\documentstyle[emulateapj]{article} 
\lefthead{Salaris et al.} 
\righthead{The distance to the LMC cluster NGC~1866 and the surrounding field} 
  
\begin{document}   
 
\title{The distance to the LMC cluster NGC~1866 and the  
surrounding field}   
  
\author{M. Salaris\altaffilmark{1},  
S. Percival\altaffilmark{1},  
E. Brocato\altaffilmark{2},   
G. Raimondo\altaffilmark{2,3}, 
A. R. Walker\altaffilmark{4}} 
    
\altaffiltext{1} 
{Astrophysics Research Institute, Liverpool John Moores   
University, Twelve Quays House, Egerton Wharf, Birkenhead CH41 1LD, UK; 
ms, smp@astro.livjm.ac.uk } 
 
\altaffiltext{2} 
{INAF - Osservatorio Astronomico di Collurania, Via M. Maggini,    
I-64100 Teramo, Italy;brocato, raimondo@te.astro.it }   
 
\altaffiltext{3} 
{Astronomia-Dipartimento di Fisica, Universit\`a La Sapienza,  P.le A.  
Moro 2, I-00185 Roma, Italy}  
 
\altaffiltext{4} {Cerro Tololo Inter-American Observatory, National 
Optical Astronomy Observatory, Casilla 603, La Serena, Chile; 
awalker@noao.edu. NOAO is operated by AURA Inc., under cooperative 
agreement with the National Science Foundation.} 
  
\begin{abstract} 
 
We use the Main Sequence stars in the LMC cluster NGC~1866 and of Red 
Clump stars in the local field to obtain two independent estimates of 
the LMC distance.  We apply an empirical Main Sequence-fitting 
technique based on a large sample of subdwarfs with accurate {\sl 
Hipparcos} parallaxes in order to estimate the cluster distance 
modulus, and the multicolor Red Clump method to derive distance and 
reddening of the LMC field.  We find that the Main Sequence-fitting 
and the Red Clump distance moduli are in significant disagreement; 
NGC~1866 distance is equal to $\rm (m-M)_{0,NGC~1866}=18.33\pm$0.08 
(consistent with a previous estimate using the same data and 
theoretical Main Sequence isochrones), while the field stars provide 
$\rm (m-M)_{0,field}=18.53\pm$0.07. This difference reflects the more 
general dichotomy in the LMC distance estimates found in the 
literature.  Various possible causes for this disagreement are 
explored, with particular attention paid to the still uncertain 
metallicity of the cluster and the star formation history of the field 
stars. 
\end{abstract} 
 
\keywords{galaxies: clusters: individual(NGC~1866) -- 
galaxies: distances and redshifts -- galaxies: individual (Large 
Magellanic Cloud) -- galaxies: stellar content} 
   
\section{Introduction}   
   
The distance to the Large Magellanic Cloud (LMC) is   
the cornerstone of the extragalactic distance scale, owing to   
the fact that the zero point of both the   
Cepheid and Type Ia supernova distances is tied to the LMC   
distance. Unfortunately, existing determinations of this   
fundamental quantity show a large range of values  
(Benedict et al.~2002), which can be schematically clustered
around $\rm (m-M)_{0,LMC}\sim$18.25--18.35 (short distance scale) and   
$\rm (m-M)_{0,LMC}\sim$18.50--18.60 (long distance scale).   
We recall that the $HST$ extragalactic distance scale   
project has determined a value for the Hubble constant   
$H_0=72\pm 3(\rm random)\pm 7(\rm systematic$) by assuming   
$\rm (m-M)_{0,LMC}=18.50\pm$0.10 (Freedman et al.~2001).   
   
This dichotomy is best illustrated by the recent results by Walker et   
al.~(2001, hereinafter W01),   
Alves et al.~(2002) and   
Salaris \& Girardi~(2002, hereinafter SG02).   
W01 have determined the distance to the LMC cluster NGC~1866 by using   
the well established Main Sequence-fitting (MS-fitting)   
technique; NGC~1866 is a well populated young cluster located about   
$4^0$ north of the center of the LMC, in a region with low extinction.   
Assuming that the cluster lies in the LMC plane the geometrical   
correction to the LMC centre is small, amounting to $\sim -$0.02 mag   
in distance modulus. W01 MS-fitting technique was based on   
theoretical isochrones which were shown to match properly the MS of   
the Hyades corrected for the Hipparcos distance modulus;   
a NGC~1866 distance modulus $\rm (m-M)_{0}\sim 18.30-18.35$ was   
obtained, a typical example of the short distance scale.   
   
On the other hand, Alves et al.~(2002) and SG02 have obtained   
$\rm (m-M)_{0,LMC}\sim 18.50$ by using multicolor photometry   
of LMC Red Clump (RC) field stars (observed in two different fields)   
as standard candles, a distance in agreement with   
the long distance scale and the $HST$ zero point. They have both used   
the local RC Hipparcos absolute brightness, corrected by the   
appropriate population corrections for the LMC computed by Girardi \&   
Salaris~(2001, hereinafter GS01) and SG02. The use of multicolor   
photometry allows one to derive   
simultaneously both distance modulus and reddening of the observed   
population, as shown by Alves et al.~(2002).  
Moreover, Sarajedini et al.~(2002) 
derive similar values by using IR observations of the red clump in  
two LMC star clusters.
   
The present investigation aims at studying in more detail this   
discrepancy between MS-fitting and RC distances to the LMC.   
We take advantage of the fact that   
the NGC~1866 $VI$ (Johnson-Cousins)   
data published by W01 show not only the well populated   
MS of NGC~1866, but also a   
clearly defined RC of the surrounding LMC field; this   
occurrence allows us to simultaneously apply MS-fitting and RC method   
to the same photometric data for the same LMC region. The advantage   
with respect to comparing MS-fitting and RC distances from different   
investigations is that in this way   
we minimize possible systematic discrepancies arising from different   
photometric zero points, differential errors in the reddening   
estimates to the observed fields, and depth effects due to the morphology  
of the LMC.   
In particular, we have redetermined the MS-fitting distance to NGC~1866   
using a completely empirical procedure (as opposed to the   
MS-fitting based on theoretical MS models performed by W01)   
which makes use of a large sample of local subdwarfs with accurate   
parallaxes presented in Percival et al.~(2003, hereinafter P03).   
At the same time, we have applied the RC method to the field RC   
stars, by using GS01 and SG02 results.   
The comparison of these distances will provide more solid evidence   
about the consistency -- or lack of it -- of the results from   
the two methods.   
Sections 2 and 3 present the distance estimates obtained from,   
respectively, MS-fitting and RC techniques. A comparison and   
discussion of the results follow in Section~4.   
   
\section{MS-fitting distance}   
   
We have used in our analysis the WFPC2 $HST$ photometry by W01 
(see also Brocato et al. 2003 for more details).  Their  
$V$-$(V-I)$ (Johnson-Cousins) Color Magnitude Diagram (CMD) shows a  
well delineated cluster MS and the LMC field Red Giant Branch and RC  
stars. In Figure~1 we show the main line of the cluster MS as  
determined by W01, together with the surrounding field population.  
   
W01 have adopted a cluster metallicity [Fe/H]=$-0.50\pm$0.1 as derived 
from high dispersion spectroscopy of three cluster giants performed by 
Hill et al.~(2000); by fitting a theoretical isochrone with 
[Fe/H]=$-$0.5 to the cluster fiducial line (see Fig.~1) both the 
reddening E$(V-I)$=0.08$\pm$0.01 and the distance modulus $\rm 
(m-M)_{V,NGC~1866}$=18.50$\pm$0.05 are obtained. The interested reader  
can find more details concerning the adopted procedure in W01.
 
This corresponds to $\rm (m-M)_{0,NGC~1866}=18.30\pm0.05$ and   
E$(B-V)$=0.064$\pm$0.011 (we adopt  
$A_I$/E(B-V)=1.8, $A_V$/E(B-V)=3.1 from Cardelli, Clayton \& Mathis~1989);   
the E$(B-V)$ value, in particular, is in good agreement with previous   
empirical determinations (van den Bergh \& Hagen~1968,  Walker~1974).     
   
W01 isochrones were also shown to fit properly the Hyades MS with the 
Hipparcos distance, when the spectroscopic metallicity [Fe/H]=$+$0.13 
is used for the isochrones.  The reliability of this NGC~1866 distance 
rests therefore entirely on the adequacy of the scaling of W01 
isochrones with [Fe/H], in the metallicity range spanned by the Hyades 
and NGC~1866. Here we have rederived the MS-fitting distance to 
NGC~1866 by following a completely empirical procedure. We have 
considered the field dwarf sample with accurate $BVI$ (Johnson-Cousins) 
photoelectric photometry and Hipparcos parallaxes presented by P03; 
these 54 objects span a [Fe/H] 
range between $\sim -$0.5 and $\sim$+0.3, and have absolute magnitudes in 
the range between $M_V\sim $5.5 and $M_V\sim $7.5, thus they are unevolved 
zero-age MS stars. Their CMD location is therefore totally insensitive 
to an age difference between NGC~1866 (age of the order of 100 Myr) 
and the local field stars, whose average age is higher than the 
cluster one, being probably of the order of a few Gyr (see, e.g., the 
discussion by Stello \& Nissen~2001). 

When determining the MS-fitting distance to a cluster, a template MS is 
constructed from the field dwarf sample by applying color shifts to the
individual stars, to account for the differences in metallicity between
the field stars and the cluster.  The procedure used to calculate the
magnitude of these metallicity dependent color shifts employs a purely 
empirical method fully described in P03, to which we refer the interested
reader for full details.  The basic method relies on first establishing
that the shape of the MS is insensitive to [Fe/H] in the narrow range of 
metallicities and magnitudes we are dealing with.  Next, we determine the 
color that each field dwarf would have at a fixed magnitude of $M_{V}=6$, 
which we call $(V-I)_{M_{V=6}}$, using the slope of the Hyades
([Fe/H]=+0.13) MS as a 
reference slope. Finally, we calculate the derivative 
$\delta(V-I)_{M_{V=6}} / \delta[Fe/H]$ which, from the full field dwarf 
sample, yields a value of $\delta(V-I)_{M_{V=6}} = 0.103\delta[Fe/H]$.
Because of the unvarying shape of the MS, this derivative is appropriate
for the whole magnitude range spanned by the field dwarf sample, and hence
color shifts are applied to each star in the sample, at their observed 
magnitudes, to construct a template MS at the metallicity of the cluster.
   
By using [Fe/H]=$-0.50\pm$0.1 and E$(V-I)$=0.08$\pm$0.01  
(the result of the fit of the field dwarfs to the cluster MS is displayed  
in Fig.~1) we obtain an empirical MS-fitting distance modulus   
$\rm (m-M)_{0,NGC~1866}=18.33\pm0.08$, in agreement   
with the results by W01 based on theoretical isochrones.   
   
\section{RC stars distance}   
   
As proposed by   
Stanek \& Garnavich~(1998), a non-linear least-square fit of the function   
\begin{equation}   
N(m_\lambda) = a + b m_\lambda + c m_\lambda^2 +   
d \exp\left[-\frac{(m_\lambda^{\rm RC}-m_\lambda)^2}   
{2\sigma_{m_\lambda}^2}\right]   
\label{eq_fit}   
\end{equation}   
to the histogram of stars in the clump region per magnitude ($m_\lambda$)   
bin has provided, among others, the apparent magnitude of the RC   
$m_\lambda^{\rm RC}$, and its associated standard error, in both   
the $V$ and $I$ photometric bands.   
We adopted   
$M_{I}^{\rm RC}=-0.26\pm0.03$ and $M_{V}^{\rm RC}=0.73\pm0.03$   
for the absolute brightness of the local Hipparcos RC   
(Alves et al.~2002), together with the population corrections by   
GS01 and SG02 which take into account the expected difference between   
the absolute magnitude   
of the RC in the solar neighbourhood and the LMC field; these corrections   
add 0.20 mag ($I$-band) and 0.26 mag ($V$-band) to the apparent   
distance moduli obtained from the local RC brightness.   
These corrections have been calculated by GS01 and SG02 from a complete
population synthesis algorithm, which produces a synthetic CMD (hence a 
luminosity function for the RC stars)
for the LMC and the solar neighborood populations, 
using the theoretical stellar models by Girardi et al.~(2000).
GS01 and SG02 have employed for the LMC 
the Star Formation Rate (SFR) determined by
Holtzman et al.~(1999; their fig.~2), and the Age Metallicity
Relationship (AMR) by Pagel \& Tautvaisiene~(1998); in the case of the
solar neighborood the SFR and AMR of Rocha-Pinto et al.~(2000a,b) have
been used. More details about
these issues and the comparison of the synthetic local RC with the
$Hipparcos$ results are given in sections 2.2, 3 and 5.4 of GS01.

Following Alves et al.~(2002), after the apparent distance moduli 
have been determined,   
one then enforces the constraint that   
the distances determined simultaneously in $V$ and $I$   
must all provide the same unreddened distance; by 
assuming the same reddening law as in Section~2,   
we obtain simultaneously  
$\rm (m-M)_{0,field}=18.53\pm$0.07, and   
E$(B-V)$=0.05$\pm$0.02. Notice the very good agreement between the   
reddening derived with this procedure and the independent estimates   
for NGC~1866 reported in the previous section.   
   
\section{Discussion}   
   
Some important results emerge from the previous two sections.   
The first one is that   
empirical and theoretical MS-fitting techniques provide exactly the   
same distance to NGC~1866, thus confirming the accuracy of the   
isochrones employed by W01 and in general the reliability of   
this method.   
The second one is that the RC method applied to the field   
around the cluster estimates a reddening   
which is in good agreement with the value for the   
cluster, and   
with the mean foreground E$(B-V)$ to the LMC, which is  
0.06$\pm$0.02 mag, according to Oestreicher, Gochermann  
\& Schmidt-Kaler~(1995).   
The third result is that   
a disturbing discrepancy between the MS-fitting distance to   
NGC~1866 and the RC distance to the surrounding LMC field   
does exist.   
The difference $\Delta$ between the LMC distances   
derived from the two methods amounts to $\Delta=0.20\pm$0.10; it is   
therefore significant at 2$\sigma$ confidence level.  
Recalling that these two independent methods are often  
used individually to obtain LMC distances with small
internal errors, the consistency between the two results appears too 
marginal to be satisfactory.
 
Let's discuss now separately possible systematic errors that can bring   
in agreement the distances obtained with the two methods.  Concerning   
the RC distance, we have mentioned the derived reddening of the field   
stars, in agreement with completely independent estimates for the   
cluster, which can be claimed as an argument for the reliability of   
the RC distance. In principle, however, the reddening derived from   
the RC does not depend on the absolute values of the population   
corrections in $V$ and $I$, but only on their difference. Population   
corrections smaller by $\sim$0.1 mag in both $V$ and $I$, but still
differing by of 0.06 mag, 
would provide the same reddening and a RC distance modulus 
in agreement with the MS-fitting one within 1$\sigma$.
   
The population corrections applied to the LMC RC depend on   
the theoretical prediction of the variation of the RC mean   
brightness for simple (single-age, single-chemical composition)   
stellar populations of varying [Fe/H] and age, and on  
the estimate of the SFR and AMR for the observed LMC field   
stars. GS01 and SG02 have clearly demonstrated with various   
empirical tests how the $V$ and $I$   
brightness of the RC in Galactic open clusters spanning a range of age and   
metallicity is well reproduced by their theoretical corrections.  
GS01 have also discussed in depth the  
issue of the LMC SFR and AMR, by using determinations for different LMC 
fields by Holtzman et al.~(1999), one of   
which is very close to our observed target. It turns out   
that for all the observed fields the individual   
population corrections are very similar, within about $\pm$0.02   
mag in both the $V$ and $I$ bands, and the difference between the $I$- and   
$V$-band correction is constant. There is therefore no indication that   
our RC distance determination is affected by sizable systematic  
errors, unless the SFR and AMR adopted to model the RC are in  
serious error.   
It is perhaps interesting to note that the distance we
obtain from the RC is in good agreement with what would be
obtained by using only the $K$-band data by Alves et al.~(2002) 
for their observed LMC field; the interest of this comparison rests on
the fact that reddening effects are negligible in the $K$-band, 
and moreover the population corrections are also negligible, at least 
for the SFR and AMR employed by GS01
and SG02. 

We have also, as an experiment, made use of  
the SFR and AMR estimated by Dolphin~(2000a) for a LMC field not far 
from NGC~1866, which,  
after computing the appropriate population corrections (GS01), provide  
$\rm (m-M)_{0,field}=18.29\pm$0.07, and   
E$(B-V)$=0.13$\pm$0.02. In this case the distance modulus is in good  
agreement with the NGC~1866 one, but the reddening is higher by a  
factor of $\sim$2, an occurrence difficult to justify    
and, moreover, the morphology of the RC as obtained  
from the theoretical simulations  
does not match the observed one, as discussed by GS01.  
   
As far as the MS-fitting distance is concerned, with our empirical   
procedure we have explored the effect of varying separately the   
cluster [Fe/H] and reddening, which are both parameters to be fixed   
beforehand. By keeping the reddening value constant at   
E$(V-I)$=0.08 (E$(B-V)$=0.064), we   
obtain that one needs a cluster metallicity between   
[Fe/H]$\sim -$0.2 and [Fe/H]$\sim +$0.2   
for having the value of $\Delta$ between zero and the associated   
$\pm$1$\sigma$ error.   
If [Fe/H] is kept fixed at $-0.5\pm$0.1,   
and the reddening allowed to vary, one needs a   
reddening E$(V-I)$ of at least 0.10 mag (E$(B-V)$=0.08) for   
the distances to agree within the 1$\sigma$ error.   
If both [Fe/H] and reddening are allowed to vary independently,   
intermediate combinations of these two quantities   
can solve the discrepancy.   
 
There is a further constraint to be applied to the cluster   
reddening. In fact, the results from our empirical procedure   
have demonstrated the reliability of the   
theoretical isochrones employed by W01, at least in   
the [Fe/H] range between the Hyades metallicity and [Fe/H]$\sim -$0.5.   
Owing to the fact that a fit of the theoretical MS to the   
cluster CMD provides both distance modulus and reddening, we can use   
theory in order to further constrain the variation of [Fe/H] and   
reddening necessary to bring the two distance methods into agreement.   
We find that by changing the cluster [Fe/H] from $-$0.5 to solar,   
the derived E$(B-V)$ changes by only 0.008;   
the value of $\Delta$ decreases to within the 1$\sigma$ error   
when [Fe/H] increases to at least   
[Fe/H]$\sim -$0.2. A negligible variation of   
E$(B-V)$ with respect to the reference value of 0.064 mag is found   
when this metallicity is adopted in the fit.   
This minimum cluster metallicity is exactly the value obtained from the   
empirical MS-fitting in case of the reddening being kept fixed.   
 
Is this value for the cluster [Fe/H] acceptable in light of the   
existing empirical constraints?   
The most direct metallicity determination for the cluster is   
the high resolution spectroscopy result adopted by W01 and by us,   
namely [Fe/H]=$-0.5\pm$0.1, which is however based on only 3 giant  
stars, and therefore cannot be considered conclusive.   
A couple of other independent   
estimates based on the integrated cluster spectrum (Oliva \&   
Origlia~1998) and on Str\"omgren photometry of supergiants (Hilker,   
Richtler \& Gieren~1995) provide values in broad   
agreement with the high resolution result, albeit with much   
larger error bars   
by $\pm$0.4 and $\pm$0.18 dex, respectively. On the other hand,   
if one employs the same Hilker et al.~(1995)   
Str\"omgren photometry of NGC~1866   
supergiants, in conjunction with the   
[Fe/H]-Str\"omgren photometry calibration   
by Arellano~Ferro \& Mendoza~(1993), an approximately   
solar [Fe/H] is obtained.   
In addition, Feast~(1989) has used $BVI$ photometry of Cepheids in the   
cluster and determined their metallicity following the procedure by   
Caldwell \& Coulson~(1985), obtaining a mean abundance [Fe/H]=$-0.1\pm$0.2.    
In light of these existing uncertainties about NGC~1866 metallicity,   
new spectroscopic observations of a wide   
sample of cluster stars (for example by using FLAMES@VLT) are   
urgently needed.   
   
We want also to mention the uncertainties due to the calibration of   
the photometric measurements. We consider a possible systematic zero-point   
error on the $V$ and $I$ WFPC2 magnitudes of $\pm$0.02 mag   
(Dolphin~2000b). Such a shift in the $V$ magnitude would marginally   
affect the distance values and do not help in decreasing substantially  
the gap between the MS and RC distances. On the other hand, a   
corresponding zero-point error by 0.03 mag on the $(V-I)$ color is expected   
to produce sizable differences for the MS-fitting distances  
due to the slope of   
the MS itself. To quantitatively explore the effect of this source of   
uncertainty on our distance evaluations we derived again the distances  
with both the MS of NGC~1866 and the RC of the surrounding field;   
by considering zero-point errors of $\pm$0.02 mag on both $V$ and $I$  
magnitudes, the systematic error on the distance moduli difference   
$\Delta$ previously defined amounts to $\pm$0.19 mag.    
Clearly this may overcome the discrepancy between the two   
methods only in the case that the assumed photometric  
zero-point is in error by $-$0.02 mag. The   
comparison with ground based photometry (Walker~1995) as performed by   
W01 does not provide any significant information, since the spread   
($\sigma \sim 0.06$) is so large that it prevents for any   
realistic conclusion on the matter.   
   
As a last point, let us recall that the previous discussion relies on   
the assumption that both the cluster and the RC field stars are located   
at the same distance from us. If this is not the case, the discussed   
dichotomy would indicate that the cluster is about 5 Kpc closer to us than   
the underlying field population of the LMC. To investigate this  
possibility one will need to determine MS-fitting distances to a  
larger sample of LMC clusters, so that depth effects due to the  
spatial distribution of the clusters cancel out when a mean value of  
their distances is computed.  
     
In conclusion, we find that the dichotomy of the distance of the LMC 
largely debated in the literature arises also when the distance to one 
single cluster (NGC~1866) and the surrounding LMC field are 
compared. This may suggest that global uncertainties in the methods of 
determining distances are underestimated. NGC~1866 also contains more 
than 20 Cepheids and this make this cluster a critical benchmark to 
probe the distance of the LMC.   
Available works on Cepheids in 
NGC~1866 (Gieren et al. 2000 and references therein)  
do not solve definitely the question, given that the derived distances 
show a sizable uncertainty. Thus, specific high 
resolution spectroscopy and detailed light curves of the cluster's 
Cepheids are required.
   
\acknowledgements{This work is partially supported by MIUR-Cofin 2000,  
under the scientific project "Stellar Observables of Cosmological  
Relevance", and by MIUR-Cofin 2002 ``Stellar populations in the Local Group as a tool
to understand galaxy formation and evolution''.}

{}   
 
\begin{figure} 
\plotone{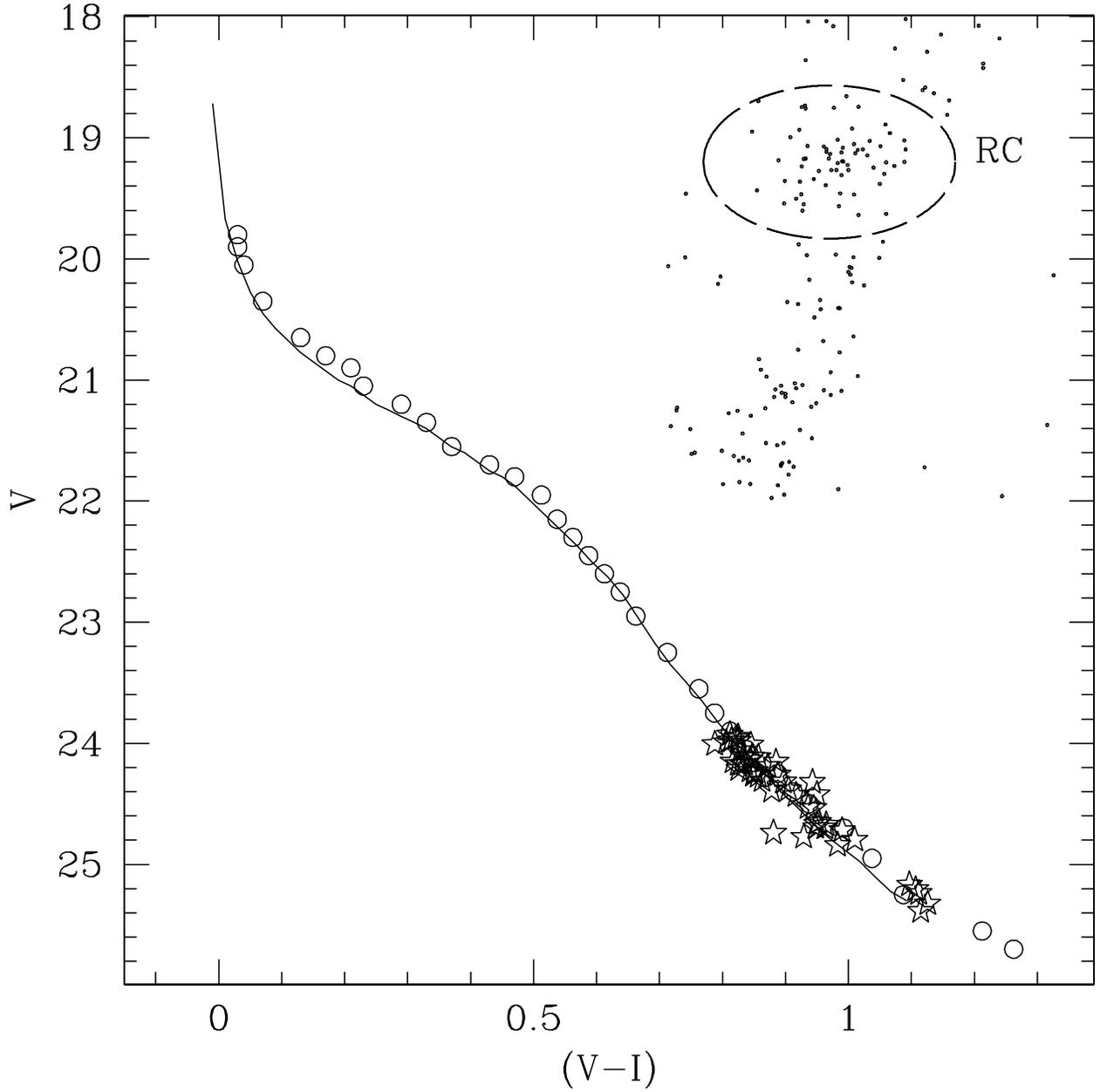} 
\figcaption{$VI$ (Johnson-Cousins) CMD of NGC~1866 MS (only the main  
sequence line is plotted as open circles), together with the Red Giant Branch   
and RC stars of the surrounding LMC field. The solid line denotes  
the best fitting isochrone with [Fe/H]=$-$0.50; star symbols show the  
sample of local subdwarfs used as standard candles,   
fitted to the cluster MS (see text for details).} 
\end{figure} 
 
\end{document}